\documentclass[10pt,twocolumn]{article} 

\usepackage{amsmath}   
\usepackage{amssymb}   
\usepackage{graphicx}  
\usepackage{natbib} 

\usepackage[hidelinks,breaklinks=true]{hyperref}

\usepackage{xurl} %
\usepackage[affil-it]{authblk} 
\usepackage{pdfpages}
\usepackage{makecell}
\usepackage[T1]{fontenc}

\graphicspath{{Fig/}}

\title{Facts are Harder Than Opinions - A Multilingual, Comparative Analysis of LLM-Based Fact-Checking Reliability}

\author[a,c]{Lorraine Saju\thanks{To whom correspondence should be addressed: \href{mailto:lorraine.saju@gesis.org}{lorraine.saju@gesis.org}}}
\author[a]{Arnim Bleier}
\author[b,c]{Jana Lasser}
\author[a,c]{Claudia Wagner}

\affil[a]{GESIS -- Leibniz Institute for the Social Sciences, Unter Sachsenhausen 6-8, 50667, North Rhine-Westphalia, Germany}
\affil[b]{University of Graz, Complexity Science Hub, Auenbruggerplatz 2, 8036, Styria, Austria}
\affil[c]{RWTH Aachen University, Chair for Computational Social Science, Templergraben 55, 52062, North Rhine-Westphalia, Germany}

\date{\today}

\begin{document}

\maketitle


\begin{abstract}
The proliferation of misinformation necessitates scalable, automated fact-checking solutions. Yet, current benchmarks often overlook multilingual and topical diversity. This paper introduces a novel, dynamically extensible data set that includes 61,514 claims in multiple languages and topics, extending existing datasets up to 2024. Through a comprehensive evaluation of five prominent Large Language Models (LLMs), including GPT-4o, GPT-3.5 Turbo, LLaMA 3.1, and Mixtral 8x7B, we identify significant performance gaps between different languages and topics. While overall GPT-4o achieves the highest accuracy, it declines to classify 43\% of claims. Across all models, factual-sounding claims are misclassified more often than opinions, revealing a key vulnerability. These findings underscore the need for caution and highlight challenges in deploying LLM-based fact-checking systems at scale.
\end{abstract}

\section*{Significance Statement}
As misinformation spreads globally, automated fact-checking solutions powered by large language models (LLMs) are gaining attention. However, current evaluations, largely limited to English political claims, may overstate LLM reliability. Using a multilingual, cross-topic dataset, we find significant weaknesses in LLM fact-checking, especially for factual statements and underrepresented languages. These results call for caution in deploying LLM-based fact-checking systems and for further work to address their vulnerabilities in diverse real-world contexts.


\section{Introduction}
The rapid spread of misinformation during critical global events, such as elections and pandemics, poses a serious threat to public trust and informed decision-making. While over 400 fact-checking organizations worldwide work to counter false claims, their capacity remains limited, with only about 110,000 fact-checks conducted annually \citep{van_damme_global_2021}. Meanwhile, generative AI technologies like ChatGPT have exacerbated the issue by automating the creation of misleading content \citep{zhou_synthetic_2023}. Compounding the problem, major platforms have scaled back content moderation \citep{vanian_tech_2023}, allowing misinformation to spread unchecked. Most notably, Meta recently announced the end of its third-party fact-checking program in the US \citep{heathera_more_2025}. Given that false information can reach peak visibility within 72 seconds and achieve 95\% of impressions within 24 hours \citep{pfeffer_half-life_2023}, manual fact-checking alone cannot keep pace. 

Recent research suggests that fact-checking interventions, such as warning labels, can be effective in reducing belief in false news, even among those skeptical of fact-checkers \citep{martel_fact-checker_2024}. However, given the scale and speed of misinformation, automated approaches are needed to complement human fact-checking efforts. Large Language Models (LLMs) such as GPT-3.5, GPT-4o, and Mixtral have demonstrated strong natural language processing capabilities \citep{ziems_can_2023}, leading to increasing interest in their potential for fact-checking. Several studies have evaluated LLMs in this context: for instance, \citet{hoes_using_2023} found that ChatGPT correctly fact-checked 72\% of 12,000 political claims from PolitiFact, while \citet{huang_fact_2023} reported that it successfully identified all 20 Alzheimer’s myths as false, with expert agreement on most explanations. Despite these promising results, LLM performance varies significantly across topics and languages, and in many cases, models refuse to classify claims or provide incorrect responses. 

A key challenge in adopting LLMs for fact-checking is determining \textit{when} they can be trusted to provide accurate classifications and \textit{when} human intervention is required. Existing studies primarily evaluate model accuracy in specific domains (e.g., U.S. politics, health misinformation) or single-language settings (predominantly English), leaving open questions about their reliability across diverse topics and languages. To address this, we systematically evaluate LLM performance on a multilingual, multi-topic dataset and examine how claim characteristics such as language, topic, and structural features influence classification outcomes. Our goal is to answer the following research question: How do different characteristics of a claim influence the accuracy of large language models in multilingual fact-checking? We investigate this by (1) analyzing the accuracy of LLM-based fact-checking across different claim characteristics, and (2) identifying factors correlated with misclassification.

Our work makes three key contributions: \\
1. \textit{A Multilingual, Multi-Topic Fact-Checked Dataset.} We compile an updated, dynamically extensible dataset of \textbf{61500} fact-checked claims from verified sources, covering \textbf{30} languages and spanning from \textbf{2007} to \textbf{2024}. This dataset allows us to evaluate LLM performance on recent, real-world claims, including those that emerged after their training data cut-off. \\
2. \textit{A Comparative Analysis of LLM Fact-Checking Accuracy.} We assess the fact-checking performance of five leading LLMs—GPT-3.5, GPT-4o, LLaMA 3.1 (8B and 70B), and Mixtral—across diverse claim types and languages. Our results show significant variation in accuracy, with GPT-4o achieving the highest performance (73.31\%) but also a critically high refusal rate (43\%), demonstrating a clear limitation for practical use. Smaller models like LLaMA 3.1 8B exhibit lower refusal rates (10.36\%) but lower accuracy (62.73\%). \\
3. \textit{An Analysis of Claim Feature Influence on LLM Fact-Checking Errors.} We analyze error rates across a range of claim features, such as language, topic, numerical data, direct quotations, and, importantly, whether the claim is presented as a fact or an opinion. Our findings reveal systematic patterns in LLM misclassifications, notably that claims presented as factual statements are significantly more prone to errors compared to those expressed as opinions. These insights into how claim characteristics impact LLM reliability provide a foundation for understanding their limitations and informing the development of targeted safeguards. 

Through our analysis of LLM performance across a diverse multilingual dataset and a range of claim characteristics, we demonstrate the nuanced nature of their reliability in fact-checking. The identified patterns of misclassification, particularly the increased error rate for factual statements, underscore the importance of understanding how claim formulation influences LLM accuracy. These insights suggest that LLMs should not be viewed as infallible, standalone fact-checkers. Instead, a more calibrated approach is needed, where their deployment is informed by a detailed understanding of their strengths and weaknesses concerning different types of claims. Our data-driven analysis contributes to a more responsible integration of AI in fact-checking by highlighting specific vulnerabilities that need to be addressed to mitigate the risks of misinformation propagation.

\section{Related Works}

\subsection{Existing Approaches to Fact-Checking}
The existing approaches to fact-checking, as delineated in the literature, span a spectrum from manual to fully automated systems. Manual fact-checking, considered the gold standard due to its potential for thorough analysis and evidence-based verdicts, involves extracting statements, constructing questions, gathering evidence, and reaching conclusions \citep{vlachos_fact_2014}. However, given the scale of information produced daily, automated or semi-automated methods are necessary.

Automated approaches typically involve a combination of machine learning, deep learning, and transformer-based methods \citep{cheung_factllama_2023}. For instance, Support Vector Machine algorithms have been used to effectively classify the veracity of claims, with \citet{hassan_detecting_2015} demonstrating a multi-tiered process that outperforms traditional methods. Deep learning approaches further refine this by directly comparing claims against evidence through complex neural networks \citep{popat_declare_2018}. Transformer-based solutions, such as those proposed by \citet{kotonya_explainable_2020}, leverage attention mechanisms to concentrate on the most pertinent evidence, which is particularly useful in the verification process.

\subsection{Nuances of Verdict Labeling in Fact-Checking}
Fact-checking has evolved from simple binary true/false labels to more nuanced multi-class systems that attempt to capture a broader spectrum of truthfulness \citep{bachenko_verification_2008, mihalcea_lie_2009}. However, there is no universal agreement on how veracity should be categorized, making consistency across fact-checking organizations a challenge. Some labels, such as mostly true or half true, introduce ambiguity, especially for composite claims that contain elements of both truth and falsehood \citep{wang_liar_2017, augenstein_multifc_2019, hanselowski_richly_2019, kotonya_explainable_2020-1, guo_survey_2022}.
Further complicating the process, claims can be technically accurate while still being misleading, creating gray areas that simple classifications cannot fully capture \citep{uscinski_epistemology_2013}. In some cases, fact-checkers provide lengthy justifications rather than a single verdict, making label standardization inherently subjective. While some researchers have explored ways to break down claims into distinct components or apply soft classification approaches \citep{pelrine_towards_2023}, these strategies introduce their own complexities. Ultimately, the challenge remains: how can diverse, often qualitative verdicts be mapped onto a standardized label set without oversimplifying the nuanced nature of misinformation?

\subsection{State of the Art in NLP Based Techniques}
Natural Language Processing (NLP) has seen a significant advance with the introduction of models like BERT, which understands text by considering context from both preceding and succeeding words (a bidirectional approach), proving to be more effective for classification tasks. In contrast, GPT-2's unidirectional approach processes text sequentially, making it useful for tasks that require sequential prediction \citep{raza_fake_2022}.

The state of the art in NLP fact-checking includes leveraging these models for their contextual understanding and generative capabilities. A prime example is the Grover framework, which closely mimics GPT-2's architecture and has been trained on extensive datasets to detect fake news \citep{zellers_defending_2020}. Despite their sophistication, these models still face challenges, such as staying current with new information and dealing with scarce labeling in real-world datasets.

\subsection{Advances and Challenges in Fact-Checking with Large Language Models}

The integration of LLMs has transformed automated fact-checking, with studies such as \citet{hoes_using_2023} showing ChatGPT’s 72\% accuracy in a zero-shot classification test of 12,784 fact-checked statements, demonstrating higher accuracy for true claims compared to false claims. LLMs like GPT-3 have been used for classifying misinformation in COVID-related tweets \citep{agresti_polimi-flatearthers_2022}, with their multilingual capabilities enabling applications across various languages.

Beyond classification, LLMs play a key role in claim detection, evidence retrieval, and verification \citep{guo_survey_2022}. Claim detection filters statements requiring verification, while LLMs assist in evidence retrieval from databases, as demonstrated by \citet{lee_language_2020}, who explored BERT's implicit knowledge for this task. However, concerns about bias and outdated knowledge prompt recommendations like \citet{cheung_factllama_2023}, suggesting external evidence retrieval mechanisms to enhance models like FACT-LLaMA. Crucially, the reliability of these models across diverse languages and topics remains a significant concern, highlighting the need for robust evaluation frameworks.

Challenges persist with biased predictions, as models trained on benchmark datasets like FEVER (Fact Extraction and VERification), which contains over 185,000 claims generated from and verified against Wikipedia, are often influenced by how a claim is worded \citep{thorne_fever_2018, schuster_towards_2019}. This has prompted the creation of adversarial claims to enhance model robustness \citep{thorne_fever20_2019}. Solutions range from new modeling approaches \citep{utama_mind_2020, thorne_elastic_2021}, to adversarial data collection \citep{eisenschlos_fool_2021}, and context-sensitive inference \citep{schuster_get_2021}.

Pre-training alignment is crucial in querying LLMs effectively, as noted by \citet{lee_language_2020}. Prompt design also influences performance, with research by \citet{gonen_demystifying_2022} and \citet{shin_autoprompt_2020} highlighting the impact of low-perplexity prompts and automated tuning. However, \citet{leidinger_language_2023} challenge the transferability of prompts across models. Biases like majority label and recency bias also affect LLM performance, as observed in \citet{zhao_calibrate_2021}.

The landscape of misinformation is further complicated by shifts in political communication styles, where leaders increasingly adopt "authentic belief speaking" that may not always align with evidential support, as observed in an analysis of US politicians on Twitter \citep{lasser_alternative_2023}. This trend highlights the challenge for LLMs in fact-checking, as the subjective nature of such statements can bias the interpretation of truth, posing difficulties in distinguishing between claims formulated in a language relating to evidence and facts, and those formulated as personal belief or intuition. This underscores the importance of developing mechanisms to identify and mitigate these biases, a key focus of our research.

The reviewed literature demonstrates a concerted effort to automate fact-checking through sophisticated NLP techniques. Despite advances in machine learning, deep learning, and transformer-based approaches, there remains a persistent challenge in transparently ensuring these models are up-to-date, context-aware, and capable of handling the nuanced and ever-changing landscape of information that necessitates fact-checking. Our review highlights 
that current evaluations are largely limited to claims from one specific topic in one language, while our work provides a multilingual, cross-topic evaluation of the fact-checking capabilities of LLMs.

\section{Methodology}

\subsection{Dataset Extension and Preparation}

To address the limitations of existing multilingual fact-checking datasets, we extended the X-Fact dataset by \citet{gupta_x-fact_2021}\footnote{MIT License, Copyright (c) 2021 Utah NLP. Full license text available at \url{https://github.com/utahnlp/x-fact/blob/main/LICENSE}} by incorporating additional claims from the \textit{ClaimReview Markup} provided via the \textit{Data Commons Feed}\footnote{\url{https://datacommons.org/factcheck/}}. ClaimReview is a structured metadata format created by Google and Duke Reporters’ Lab in 2015\footnote{See \citet{lim_better_2019} and \url{https://schema.org/ClaimReview}.}, designed to make fact-checking articles more discoverable and machine-readable. It is widely adopted by platforms such as Google, Facebook, and DuckDuckGo, and is used by fact-checking organizations to share structured information about verified claims. The \textit{Data Commons Feed} aggregates ClaimReview-annotated content from a variety of fact-checking sources across the web.

Our goal was to build an up-to-date and high-quality multilingual dataset by merging the original X-Fact dataset with cleaned and filtered ClaimReview data. This enabled us to analyze model performance on claims published both before and after the training data cut-off of the language models.

\subsubsection{Data Collection and Filtering}

The data collection and cleaning process involved four main stages:

\paragraph{Data Collection:}
We used the original X-Fact dataset (up to 2020), which included 43,483 claims. To expand this dataset with more recent content, we extracted approximately 51,000 additional claims from the Data Commons Feed, which aggregates fact-checks from ClaimReview-annotated sources across the web.

\paragraph{Source Verification and De-duplication:}
To ensure the credibility of fact-checking sources, we filtered the post-2020 data to retain only claims published by organizations recognized by the International Fact-Checking Network (IFCN)\footnote{\url{https://www.poynter.org/ifcn/}} or listed in the Duke Reporters’ Lab\footnote{\url{https://reporterslab.org/fact-checking/}}. This step reduced the set to 43,938 claims. We also ensured that there was no duplication of claims from the year 2020 between the X-Fact dataset and the Data Commons Feed.

\paragraph{Verifiability and Media Filtering:}
We followed the definition of verifiability from \citet{pelrine_towards_2023}, which requires a claim to contain sufficient textual information for its truthfulness to be assessed. To apply this filter at scale, we used GPT-3.5 and GPT-4 with custom prompts (detailed in the Supplementary Material) to evaluate each claim’s verifiability. This resulted in 33,430 verifiable claims from X-Fact and 34,065 from the post-2020 data.
We also filtered out claims tied to non-textual elements (e.g., videos or images), which are unsuitable for purely text-based fact-checking models. Using a media-focused GPT prompt, we removed an additional 5,204 claims that were strongly dependent on such content.
At this stage, we merged the cleaned X-Fact and ClaimReview-derived datasets, resulting in a unified collection of high-quality, verifiable, and text-based claims suitable for multilingual fact-checking analysis.

\paragraph{Language-Based Filtering:}
To support language-specific analysis with meaningful sample sizes, we excluded languages with fewer than 100 claims. This threshold allowed us to balance multilingual coverage with statistical reliability. The distribution of languages and their respective claim counts is provided in the supplementary material.

\begin{table*}[htbp]
\centering
\caption{Dataset Statistics}
\begin{tabular}{|l|p{12cm}|}
\hline
\textbf{Characteristic}        & \textbf{Value}                \\ \hline
Total Claim Count     & 61514                                   \\ \hline
Number of Languages   & 30                                      \\ \hline
Languages             & English, Portuguese, Spanish, German, Indonesian, Tamil, Hindi, Arabic, Turkish, Polish, Italian, Telugu, Dutch, Romanian, French, Persian, Serbian, Bengali, Georgian, Sinhala, Russian, Albanian, Norwegian, Kannada, Filipino, Hebrew, Taiwanese Mandarin, Azerbaijani, Assamese, Khmer \\ \hline
Time Span             & 2007 - 2024                             \\ \hline
Number of Topics      & 5  \\ \hline
Topic Distribution & Politics and Governance (25.19\%), Society and Culture (24.00\%), Economy and Environment (21.02\%), Health and Pandemics (16.74\%), Conflict and Security (13.05\%) \\ \hline
Claim Type Distribution & Expressed as Opinion (60.12\%), Stated as Facts (39.88\%)         \\ \hline
\end{tabular}
\label{tab:dataset_statistics}
\end{table*}

\subsubsection{Verdict Label Normalization}
To standardize the dataset for training and evaluation, we mapped diverse verdict labels from various fact-checking organizations into five unified categories: \textit{False}, \textit{Mostly False}, \textit{Partly False/Misleading}, \textit{Mostly True}, and \textit{True}. This normalization process was essential due to the heterogeneity in rating schemes and label languages used by different sources.

We began with the label mappings provided in the original X-Fact repository\footnote{\url{https://github.com/utahnlp/x-fact/tree/main/data/x-fact/label_maps}}, which included verdict translations and harmonizations for major fact-checkers such as PolitiFact and Chequeado. These mappings were semi-automatically aligned using synonym detection and manual translation, following the approach described by \citet{gupta_x-fact_2021}. For example, “pants on fire!” was mapped to \textit{False}, and “half-true” to \textit{Partly False/Misleading}.

To accommodate new verdict labels encountered in the expanded ClaimReview dataset, we manually curated additional mappings. This included verdicts in multiple languages and organization-specific phrasing, which we normalized to match our five-label scheme. Where necessary, we consulted the source website’s rating definitions to ensure alignment with the original intent of the label.

\subsubsection{Annotation Process}
To enrich the dataset, we applied three types of annotations using GPT-3.5 Turbo for automated extraction. Each annotation task was designed to provide deeper insight into claim characteristics relevant to the fact-checking process.

\begin{enumerate}
    \item \textbf{Topic Extraction:} 
    Traditional topic modeling techniques, such as Latent Dirichlet Allocation (LDA), often struggle with semantic coherence and produce overlapping or ambiguous topic groupings. Recent work has shown that LLMs can serve as more flexible and interpretable alternatives for topic modeling, capable of generating coherent topic titles and refining or merging categories based on semantic understanding and minimal guidance \citep{mu_large_2024}.

    Motivated by these advances, we employed a semi-supervised approach using GPT-3.5 Turbo to annotate topics. We began by prompting the model to assign topics to a random subset of claims without providing predefined options. The LLM-generated topics were then manually reviewed, and semantically similar categories were iteratively merged into five broad groups: \textit{Health and Pandemics}, \textit{Politics and Governance}, \textit{Society and Culture}, \textit{Economy and Environment}, and \textit{Conflict and Security}. This process leveraged the model’s capacity for clustering thematically related content and provided a consistent topical framework for the full dataset.

    Manual validation of 50 annotated claims confirmed that the topic labels were appropriate in all cases.

    \item \textbf{Claim Type Identification:} 
    Beyond topic modeling, LLMs have shown strong capabilities in psychological and discourse-level text classification. Recent studies demonstrate that models like GPT outperform traditional approaches in identifying nuanced constructs such as sentiment, emotion, and moral framing, and are also capable of distinguishing between factual and opinion-based content \citep{rathje_gpt_2024, broekens_fine-grained_2023}. These advances position LLMs as effective tools for analyzing subjective or ambiguous claims in a fact-checking context.

    Building on this, we used GPT-3.5 Turbo to classify each claim as either \textit{factual} or \textit{opinion}-based, following the methodology of \citet{broekens_fine-grained_2023}. A claim was labeled \textit{factual} if it was presented as a verifiable statement, and \textit{opinion} if it expressed personal beliefs, evaluations, or subjective interpretations.

    To validate the reliability of this classification, we manually reviewed a random sample of 50 annotated claims. The annotations were accurate in 48 out of 50 cases, confirming the model's strong performance in distinguishing between factual and opinionated language.

    \item \textbf{Further key features:} 
    Additional structural features were annotated for each claim: \textit{Numerical Claims}, \textit{Quotes}, \textit{Position Statements}, and \textit{Entity/Event Properties}. These features were first identified through manual inspection of a representative subset of claims, and then annotated via LLM prompting. To ensure accuracy, we manually reviewed a random sample of 50 claims to confirm that the feature annotations were consistent with human expectations. The schema and examples were iteratively refined during this review process to align with human annotation standards.

\end{enumerate}

This newly created, \textit{dynamically extensible} dataset, called FactSpan,  (detailed in Table \ref{tab:dataset_statistics}) covering 30 languages and around 61,500 claims, is made publicly available\footnote{\url{https://zenodo.org/records/15084388}} , along with our data processing code\footnote{\url{https://github.com/lorraine-dev/FactSpan}}, to facilitate continuous updates and future research. This is the first dataset of its kind, to our knowledge, that allows for easy updating to include current fact checking data. The creation of this dataset represents a significant advancement in the field of multilingual fact-checking. By providing a large, up-to-date, and dynamically extensible resource, this dataset empowers researchers and practitioners to develop and evaluate more robust and reliable LLM-based fact-checking systems.

\subsection{Model Selection and Fact-Checking}

For this study, we selected five widely available LLMs to evaluate their performance in automated fact-checking tasks, covering both closed-source and open-source options. From the closed-source models, we tested OpenAI's more affordable GPT-3.5, as well as the more advanced GPT-4o, which claims to provide enhanced accuracy at a higher cost. To explore open-source alternatives, we chose LLaMA 3.1 (70B) and Mixtral (8x7B), both of which are known for their strong performance in multilingual tasks \citep{dubey2024llama, jiang_mixtral_2024}. Additionally, to assess the performance of smaller open-source models, we included LLaMA 3.1 (8B) in our experiments.

Although our dataset contained five standardized verdict labels indicating varying degrees of veracity, we simplified the fact-checking task by asking the LLMs to perform binary classification: \textit{True} or \textit{False}. To evaluate model performance, we mapped \textit{False}, \textit{Mostly False}, and \textit{Partly False/Misleading} to \textit{False}, and \textit{Mostly True} and \textit{True} to \textit{True}. Results were compared against manual fact-checker labels using accuracy, precision, recall, and F1-score.

We designed a base prompt as a starting point for each model, adjusting parameters such as temperature (set at or near 0) to ensure deterministic and reproducible outputs for fact-checking tasks. Each model was fed batches of claims, with batch sizes optimized for each model's context length and API rate limits. Specifically, LLaMA 3.1 8B was hosted locally on the institution's GPU server, while LLaMA 3.1 (70B) and Mixtral (8x7B) were accessed via the Groq API. All OpenAI models were accessed via their respective APIs.

\begin{quote}
Base Prompt: "Fact check the following statements. Use labels ’true’ or ’false’. If a verdict cannot be reached, use ’no verdict’. The output should only contain the numbered labels corresponding to each statement."
\end{quote}
The base prompt was fine-tuned iteratively for each model, modifying instructions and structure to optimize performance during testing. This prompt optimization allowed for fair comparisons between the models, while ensuring that each was tested under conditions suited to its architecture and capabilities. The specific model versions, the prompts used, and information about processing speed are provided in the Supplementary Material.

\subsubsection{Language Resource Taxonomy}
\label{subsec:language_taxonomy}

To better understand the impact of language resource availability on the performance of LLMs in multilingual fact-checking, we incorporated the language resource taxonomy proposed by \citet{joshi_state_2020}. This taxonomy categorizes languages into six classes (0-5) based on their digital resource richness.

\begin{itemize}
    \item \textbf{Class 0 (The Left-Behinds):} Languages with exceptionally limited resources.
    \item \textbf{Class 1 (The Scraping-Bys):} Languages with some unlabeled data, but limited labeled data.
    \item \textbf{Class 2 (The Hopefuls):} Languages with a small set of labeled datasets.
    \item \textbf{Class 3 (The Rising Stars):} Languages with strong web presence and unsupervised pre-training potential.
    \item \textbf{Class 4 (The Underdogs):} Languages with abundant unlabeled data but less labeled data.
    \item \textbf{Class 5 (The Winners):} Rich-resource languages with abundant labeled and unlabeled data.
\end{itemize}

The specific mapping of languages in our dataset to these classes, including details on the mapping process and any language-specific adjustments, is provided in the Supplementary Material. This mapping allowed us to analyze how LLM performance varies between languages with different levels of digital resource availability.

\subsection{Analysis of Claim Features related to LLM Accuracy}

To understand the factors affecting the reliability of LLM-based fact-checking, we conducted a detailed analysis of how different claim features correlate with LLM accuracy. Our methodology focused on identifying systematic patterns in LLM errors across a range of claim characteristics.

\paragraph{Feature-Wise Error Rate Analysis} 
First, we conduct a simple error rate analysis to examine how different claim attributes relate to misclassification rates. We compute error rates for each feature independently—such as claim type (fact vs. opinion), topic, language, and the presence of features like Quotes and Numerical Data—to identify high-risk claim categories. Additionally, we extend this analysis to multi-feature combinations, uncovering interactions between attributes that may contribute to LLM misclassifications. To ensure statistical robustness, feature combinations with low data counts were excluded using a minimum threshold which varied depending on the feature distribution. This ensured that error rate estimates were not skewed by small sample sizes. 

\paragraph{Error Rate Analysis for Factual vs. Opinion Claims}
Recognizing the nuances in modern communication, where subjective beliefs are often presented in a manner that can be mistaken for factual assertions, we specifically examined the misclassification rates for claims categorized as factual versus those expressed as opinions for each model. This comparative analysis across the five different LLMs aimed to identify potential differences in how these models process and evaluate different style of claims.

Our approach prioritizes understanding the specific claim characteristics that contribute to LLM fact-checking errors. By performing a comprehensive feature-wise analysis, with a particular emphasis on the distinction between factual and opinion-based claims across multiple LLMs, we aim to provide insights into the models' strengths and weaknesses. This detailed examination lays the groundwork for developing more informed strategies for deploying LLMs in fact-checking workflows.

\section{Results}
\subsection{Fact-Checking Performance of LLMs}

\begin{figure*}[!p]%
\centering
\includegraphics[width=0.9\textwidth]{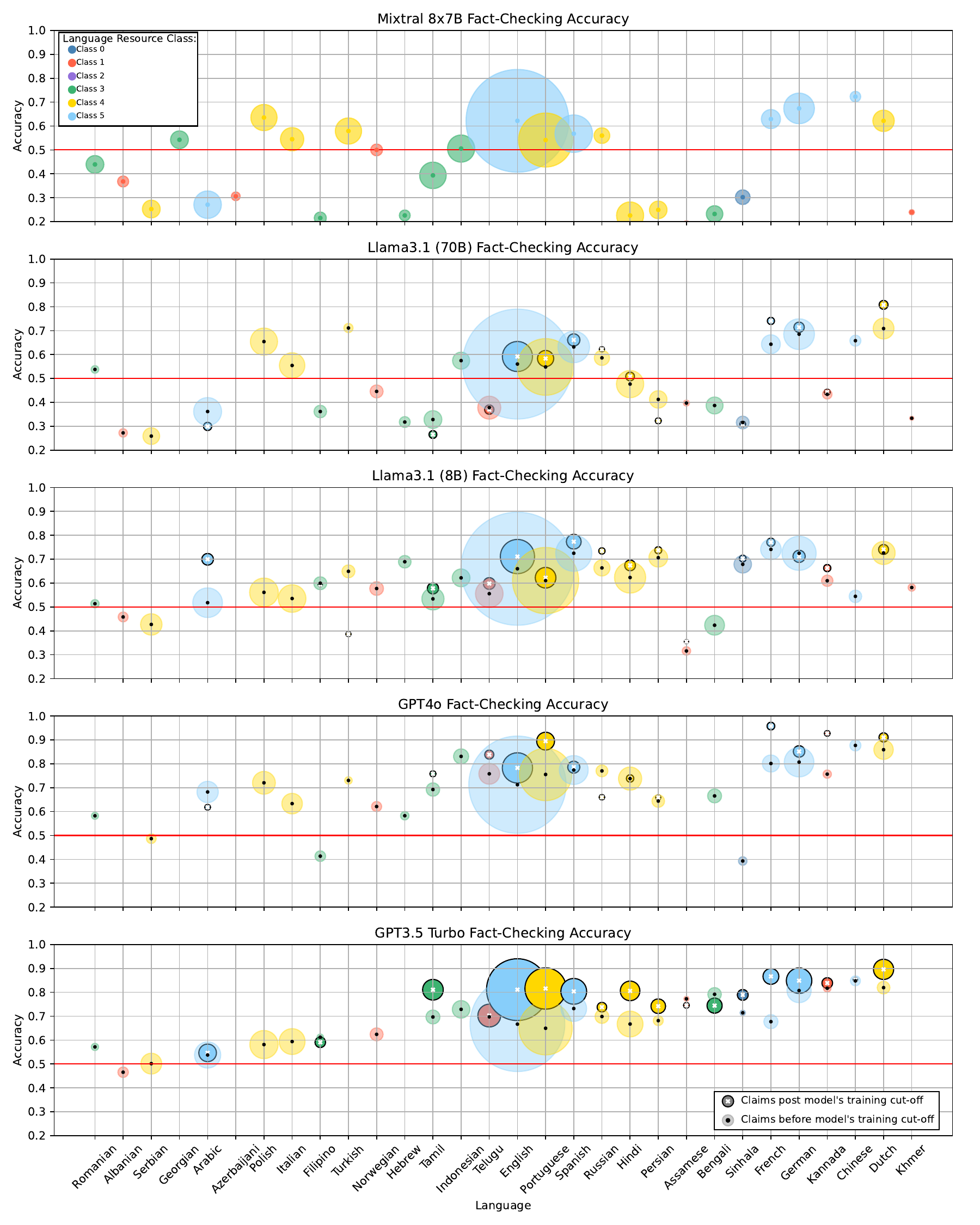} 
\caption{\footnotesize Fact-checking accuracy of five LLMs (Mixtral 8x7B, LLaMA 3.1 (70B and 8B), GPT-4o, and GPT-3.5 Turbo) across 61,514 claims in multiple languages. Each subplot represents one model, with the x-axis showing different languages and the y-axis representing the fact-checking accuracy. For each language, we display the model’s performance on claims created before the model’s training cut-off (black dot) and after the cut-off (white cross). The size of the circle around each marker corresponds to the number of claims available for that language. The red line at 0.5 indicates chance-level performance, as fact-checking was framed as a binary classification task. Circle color denotes language resource class, based on the \cite{joshi_state_2020} taxonomy, with Class 5 being high-resource languages and Class 0 low-resource.\\
GPT-3.5 Turbo and GPT-4o exhibit strong generalization beyond the official training cut-off dates, often outperforming their own pre-cutoff performance, especially in high-resource languages. The LLaMA 3.1 models also show slight improvements post-cutoff, though the gains are more modest. Mixtral's cut-off date is not publicly known, so pre/post distinctions are not shown for that model. The figure also illustrates large performance variability across languages and models, emphasizing the role of both language resources and model scale.}
\label{allmodels_results}
\end{figure*}

We conducted a comprehensive evaluation of five LLMs for automated fact-checking across multiple languages. Each model was tasked with fact-checking 61,514 claims. The models classified claims as either \textit{True}, \textit{False}, or returned \textit{No verdict} when a conclusion could not be reached. In this section, we present the performance of each model, focusing on overall accuracy, the percentage of claims for which no verdict was reached, and how each model performed on claims made before and after the model’s respective training data cut-off points. The performance of each model is illustrated in Figure \ref{allmodels_results}, which displays the accuracy by language and the distinction between pre- and post-cutoff claims where applicable. The color of each point in Figure \ref{allmodels_results}, indicates the language resource class, as defined by \citet{joshi_state_2020} and detailed in Section \ref{subsec:language_taxonomy}, where class 5 is the most digitally available, and class 0 the least.

All models processed the same 61,514 claims, but their performance varied significantly. \textbf{Mixtral 8x7B} had the lowest overall accuracy of 53.41\%, with a high \textit{No verdict} rate of 36.68\%, highlighting it's unreliability for fact-checking as its accuracy barely exceeds chance. \textbf{LLaMA 3.1 70B} had 25.57\% \textit{No verdicts} and 54.55\% accuracy. The smaller \textbf{LLaMA 3.1 8B} surprisingly performed better, with the lowest \textit{No verdict} rate (10.36\%) among all models and the highest accuracy of 63.82\% among the open-source models. Among the closed-source models, \textbf{GPT-3.5} achieved 69.44\% accuracy with 16.57\% \textit{No verdicts}, while \textbf{GPT-4o} had the highest accuracy of all models at 73.31\% but also the highest \textit{No verdict} rate (43.02\%). More details about the No verdict percentages for the models per language is given in the Supplementary Material.

\begin{figure*}[!t]
\centering
\includegraphics[width=0.9\textwidth]{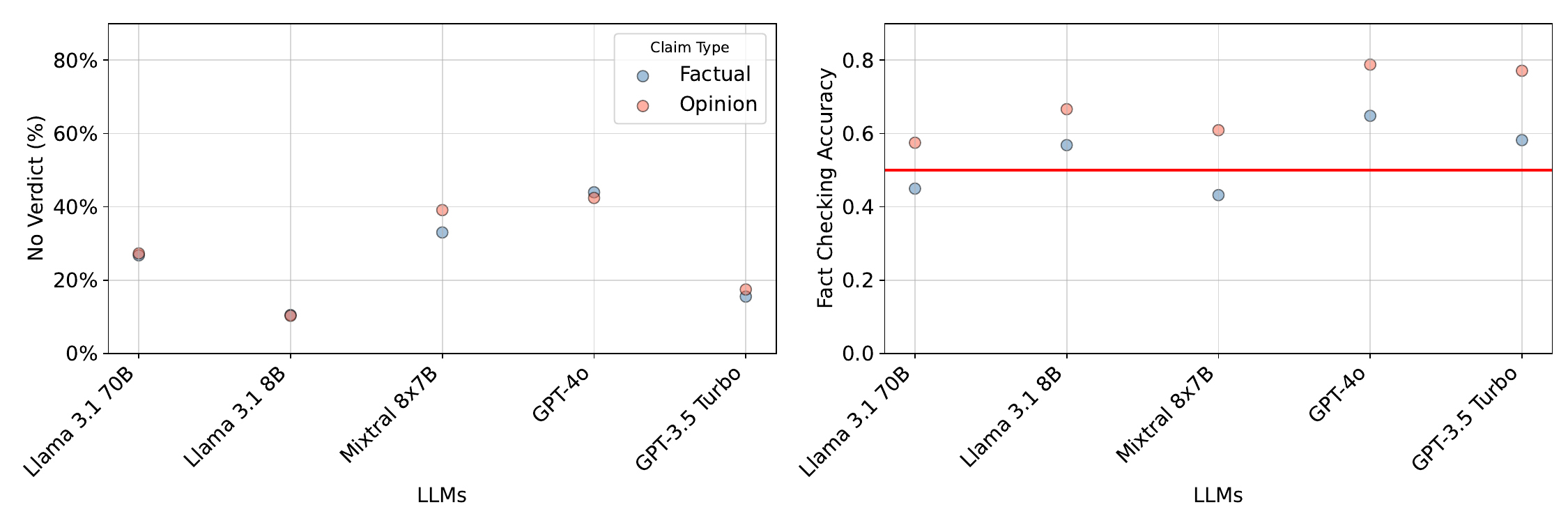}
\caption{\footnotesize Performance of Five LLMs on Factual vs. Opinion Claims: (Left) 'No Verdict' Percentage and (Right) Fact-Checking Accuracy, with a red line at .5 indicating chance-level performance for the accuracy subplot. Points are colored by claim type, allowing for a direct comparison of model behavior across these two categories.}
\label{fig:combined_scatter}
\end{figure*}


\subsection{Analysis of Claim Features related to LLM Accuracy}
To gain deeper insights into the factors influencing LLM fact-checking accuracy, we conducted a detailed analysis of how different claim features correlate with the performance of the models. Our initial analysis focused on fact-checking results from GPT-3.5 Turbo to identify potential patterns.
\paragraph{Feature-Wise Error Rate Analysis (GPT-3.5 Turbo)} The error rate analysis for the LLM revealed several interesting trends:
\begin{itemize}
\item \textbf{Claim Type:} Claims stated as facts had a significantly higher error rate (\textbf{41.2\%}) than claims expressed as opinions (\textbf{21.3\%}), suggesting that GPT-3.5 struggles more when verifying objective statements.
\item \textbf{Claim Topics:} Claims related to \textbf{Economy and Environment} had the highest error rate (\textbf{37.1\%}), whereas claims about \textbf{Health and Pandemics} had a lower error rate (\textbf{21.9\%}).
\item \textbf{Language-Specific Performance:} The highest error rates were observed in \textbf{Serbian} (\textbf{49.9\%}), \textbf{Arabic} (\textbf{46\%}), and \textbf{Polish} (\textbf{41.9\%}). In contrast, \textbf{Dutch} (\textbf{12.5\%}) and \textbf{German} (\textbf{17.1\%}) had the lowest error rates, suggesting that LLM performance varies significantly by language.
\end{itemize}

\paragraph{Multi-Feature Error Analysis}
Examining the interaction between multiple attributes, we found that error rates were particularly high for claims that combined multiple challenging factors. For instance:

\begin{itemize}
\item \textbf{Claims stated as facts about Economy and Environment in Serbian (pre-2019)} exhibited the highest error rate (\textbf{67.6\%}).
\item \textbf{Claims expressed as opinions about Health and Pandemics in German (post-2019)} had a much lower error rate (\textbf{5.2\%}).
\end{itemize}

\paragraph{Logistic Regression Results}
To examine the combined effect of multiple claim attributes on the fact-checking accuracy of a LLM, we conducted a logistic regression analysis. Key findings include:

\begin{itemize}
\item \textbf{Claim Age:} The likelihood of an error decreased by \textbf{9.8\% per year} (odds ratio, OR = \textbf{1.098}), indicating that GPT-3.5 performed better on more recent claims.
\item \textbf{Language Impact:} Compared to English, claims in \textbf{Arabic} (OR = \textbf{0.43}), \textbf{Italian} (OR = \textbf{0.50}), and \textbf{Polish} (OR = \textbf{0.58}) had significantly higher odds of being misclassified.
\item \textbf{Claim Type:} Claims formulated as facts were significantly harder to classify accurately, with opinion-based claims having \textbf{2.7 times lower odds of error} (OR = \textbf{2.70}).
\item \textbf{Complex Labels:} Claims labeled as \textbf{Partly True/Misleading} were more likely to be misclassified than those labeled as \textbf{False} (OR = \textbf{0.50}).
\end{itemize}

While the presence of numerical data did not significantly impact the error rate in isolation, presence of features such as \textbf{Entity/Event Properties} and \textbf{Position Statements} emerged as important predictors of classification difficulty.

Our analysis highlights key areas where GPT-3.5's fact-checking performance varies:

\begin{itemize}
\item \textbf{Facts are harder than opinions:} Claims articulated as factual statements were significantly more challenging for GPT-3.5.
\item \textbf{Certain languages pose greater difficulty:} Fact-checking accuracy was notably lower in Serbian, Arabic, and Polish than in other languages.
\item \textbf{Claim complexity matters:} More nuanced claim labels (e.g., "Partly True/Misleading") led to higher misclassification rates.
\end{itemize}

\paragraph{Claim Type and LLM Accuracy Across Models}
We conducted a focused analysis on the impact of claim type (factual vs. opinion) across all five evaluated LLMs. As illustrated in Figure \ref{fig:combined_scatter}, the accuracy scores revealed a consistent and substantial difference. For all five LLMs, the accuracy for claims stated as facts was consistently and significantly lower than the accuracy for claims expressed as opinions. Specifically:

\begin{itemize}
    \item LLaMA 3.1 70B achieved an accuracy of 44.97\% on factual claims compared to 57.48\% on opinions.
    \item LLaMA 3.1 8B showed an accuracy of 56.83\% on factual claims versus 66.64\% on opinions.
    \item Mixtral 8x7B's accuracy was 43.20\% for factual claims and 60.90\% for opinions.
    \item GPT-3.5 Turbo achieved 58.19\% accuracy on factual claims and 77.11\% on opinions.
    \item GPT-4o demonstrated 64.83\% accuracy on factual claims and 78.79\% on opinions.
\end{itemize}

Notably, the 'No Verdict \%' did not show a clear or consistent relationship with claim type across the models (as shown in Figure \ref{fig:combined_scatter}). However, the consistent and substantial accuracy gap strongly suggests a potential inherent challenge in processing and verifying factual statements compared to subjective expressions across a range of LLM architectures. These findings underscore the importance of considering claim formulation when evaluating the reliability of LLMs in fact-checking.

\subsection{Pre-Cutoff vs. Post-Cutoff Performance}

We compared each model's performance on claims before and after their respective training data cut-off points. One would expect that models perform better on pre-cutoff data since they might have seen this data already. However, Figure \ref{allmodels_results} shows that for most models and languages this is not the case. Figure \ref{allmodels_results} illustrates the fact-checking accuracy of each model across different languages, with marker size indicating the volume of claims per language. Specifically, the figure shows the model's performance on claims created \textit{before} the model’s training cut-off (black dot) and \textit{after} the cut-off (white cross), allowing for a direct comparison of pre- and post-cutoff accuracy. This visualization highlights the fluctuation of model accuracy based on language and claim volume, and shows that the closed-source models generally outperform the open-source models, especially on post-cutoff data. GPT-3.5 Turbo has a training data cut-off of September 2021, GPT-4o October 2023, and LLaMA 3.1 December 2023, while information on Mixtral’s cut-off date is unavailable. GPT-3.5 and GPT-4o showed strong adaptability to post-cutoff data, with GPT-3.5 achieving \textbf{79.75\%} post-cutoff accuracy (compared to \textbf{66.05\%} pre-cutoff) and GPT-4o reaching \textbf{80.45\%} post-cutoff (versus \textbf{72.57\%} pre-cutoff), indicating robust generalization. In contrast, the open-source models, LLaMA 3.1 70B and 8B, had more modest improvements, with LLaMA 3.1 8B improving from \textbf{63.33\%} pre-cutoff to \textbf{68.55\%} post-cutoff, while LLaMA 3.1 70B remained largely consistent around \textbf{55\%}.

Performance differences between closed-source and open-source models may be explained, at least in part, by the fact that published open-source models are typically not further tweaked with human feedback after their cut-off date. In contrast, for closed-source models, the inclusion of data beyond the official cut-off date has been documented \citep{openai_model_release_notes}, making additional fine-tuning of these models at least theoretically possible. Nevertheless, further investigations are needed to fully understand the mechanisms underlying these performance differences. It is also important to acknowledge the limitations of using claim date as a direct proxy for event knowledge. A claim made after the model's cut-off could refer to an event that occurred before the cut-off date, and therefore the model has already seen that event. Moreover, recent research \citep{tai_genai_2025} suggests that LLMs may rely heavily on linguistic features and "hard" criteria, such as language formality and level of detail, rather than a genuine understanding of veracity. This could also contribute to the models' ability to accurately classify claims about events occurring after their training cut-off. A more detailed analysis of this observation is provided in the Supplementary Material.

\section{Discussion and Conclusions}

This research addressed two key objectives: the creation of a novel, dynamically extensible multilingual dataset and a comprehensive evaluation of LLM performance in fact-checking.

The dynamically extensible dataset, a significant contribution of this work, is designed to be continuously expanded by researchers using the accompanying update script provided in the repository. Unlike static datasets, it enables the incorporation of new claims at any time, ensuring ongoing relevance in the rapidly evolving landscape of misinformation. This flexibility supports longitudinal studies and adaptation to emerging topics or languages. Using this dataset, we conducted a comprehensive evaluation of LLM performance across diverse languages and claim types, uncovering substantial performance variation. Notably, GPT-4o achieved the highest accuracy but also showed a high refusal rate, highlighting a trade-off between precision and coverage.

One of the most striking findings was the performance gap between claims expressed as opinions and those presented as facts. LLMs performed better on the former, likely due to their reliance on linguistic heuristics such as tone, formality, and plausibility rather than engagement with factual knowledge. In contrast, claims presented as facts often require more specific reasoning or external world knowledge, which LLMs might not always consistently apply. \textbf{This discrepancy may lead to higher rates of misclassification, particularly when claims are framed to resemble factual statements without being verifiably accurate.} The study highlights the need for LLMs to not only detect linguistic signals but also reason through content grounded in verifiable information. For future research it would be interesting to test the fact-checking reliability of LLMs on opinion and factual claims experimentally.

Further analysis of model behavior on pre- and post-training cutoff claims revealed that LLMs can perform surprisingly well even on claims occurring after their advertised training data window. This adaptability leaves open the possibility that the models are not solely dependent on factual memory but may also be leveraging linguistic and semantic heuristics—such as plausibility, language formality, and narrative structure—to evaluate claim veracity. While this allows LLMs to classify some post-cutoff claims with reasonable accuracy, it also underscores a critical limitation: accuracy may not reflect genuine factual understanding. This raises important concerns about the real-world reliability of these systems, as high performance on benchmark datasets might not translate to dependable fact-checking in dynamic, real-world scenarios. This finding aligns with recent work \citep{tai_genai_2025}, which argues that LLMs often substitute linguistic fluency for epistemic accuracy.

Finally, our analysis revealed that misclassification rates were influenced by multiple dimensions beyond claim type, including language and topic. For instance, performance varied considerably in languages like Serbian and Arabic and for complex thematic areas. These findings emphasize that LLM fact-checking performance cannot be evaluated solely based on language proficiency or model size but must account for interaction effects between linguistic, topical, and structural features of claims.

Together, these findings underscore the importance of critically evaluating how LLMs interpret and classify claims. As LLMs are increasingly integrated into real-world fact-checking workflows, understanding their limitations, including their reliance on surface-level cues and linguistic heuristics is essential. Ensuring that models can distinguish between claims that sound factual and those that are factual will be key to improving trust and reliability in automated fact-checking systems.

\subsection{Challenges, Limitations, and Directions for Future Research}

Several limitations of this study warrant attention. First, while a claim’s publication date provides a rough proxy for event novelty, claims about earlier events can still surface after a model’s training, complicating interpretations of post-cutoff performance. Second, the dataset, though multilingual, remains skewed toward languages and topics with more active fact-checking communities.

Future work should focus on enriching the dataset with claims in low-resource languages and across a broader range of claim types and difficulty levels. In addition, better tools for assessing LLM reasoning processes—beyond simple accuracy measures—are needed to uncover whether models are genuinely verifying facts or merely applying linguistic heuristics.

Furthermore, while this study focused on text-based claims, the increasing prevalence of misinformation in non-textual formats necessitates the development of multi-modal fact-checking models that can address claims embedded in images, videos, and other media. Integration of external knowledge sources and multi-modal reasoning capabilities could significantly enhance the robustness of future fact-checking systems.

Finally, although GPT-3.5 Turbo offered a scalable and effective approach for automating claim annotations, the annotations may still reflect biases present in the model’s training data. This risk is particularly relevant for subjective or culturally sensitive topics, where model outputs may inadvertently reflect dominant narratives or regional biases. While we conducted manual validation to confirm high accuracy for core annotation tasks (e.g., topic and claim type), a more systematic error analysis or cross-cultural evaluation would strengthen future work. Another limitation lies in our prompt design. While we iteratively refined the prompt to optimize performance, we acknowledge the possibility of prompt bias. To minimize this risk, we kept the prompt wording as neutral as possible and ensured consistency across all models. Nonetheless, subtle differences in how each model interprets prompts may still affect outcomes.

\subsection{Contributions and Final Thoughts}

This research makes two primary contributions: the creation of a dynamically extensible multilingual fact-checking dataset, and a detailed evaluation of current LLM fact-checking capabilities across models and languages, while also analysing claim features related to LLM accuracy.

By systematically examining how claim type, event recency, and language affect LLM performance, this study sheds light on critical limitations and opportunities for improvement. As misinformation continues to evolve, the development of scalable, accurate, and cautious fact-checking systems remains essential. Through continued refinement of both LLM architectures and supporting evaluation methods, we can move closer to reliable, responsible fact-checking solutions for the global information ecosystem.

\section{Data availability}
The dataset underlying this article is available on Zenodo at \url{https://zenodo.org/records/15084388} and can be accessed via DOI: 10.5281/zenodo.15084388. The dataset can be dynamically extended using the code provided at \url{https://github.com/lorraine-dev/FactSpan}.

\section{Funding}
This work was partially supported by the German Research Foundation (DFG, project no. 504226141).

\bibliographystyle{plainnat}
\bibliography{reference.bib}

\clearpage
\onecolumn
\section*{Supplementary Materials} 

\section*{Methods}

\subsection*{Dataset Extension and Preparation}

\subsubsection*{Prompts used in the paper} 
\leavevmode

To ensure the claims in our dataset were suitable for text-based fact-checking, we implemented a verifiability check, as described in Pelrine et al.\cite{pelrine_towards_2023}. This process aimed to identify claims that contained sufficient textual information for independent assessment, excluding those that were truncated or required non-textual context (e.g., images, videos).

The following prompt was used to assess the verifiability of each claim:

\begin{quote}
Determine if each claim is fact-checkable. Label each as 'Verifiable' or 'Unverifiable'. A claim is unverifiable if it is truncated or requires non-textual context (like an image or video). Use your pre-training knowledge to understand non-English claims based on linguistic and cultural context. Here are examples to guide your judgment:

Input:
1. "A recent study found that eating two apples a day can drastically reduce your risk of heart disease." - This claim mentions a study with specific outcomes, making it verifiable.

2. "Watch this video to see the shocking truth about climate change." - This claim requires viewing a video, which is non-textual, making it unverifiable.

Output:
'1. Verifiable', '2. Unverifiable'
\end{quote}

To optimize computational costs while maintaining accuracy, we employed a two-pass approach using a combination of GPT-3.5-turbo and GPT-4.

\textit{First Pass: GPT-3.5-turbo}: Initially, we applied the verifiability check to all claims using GPT-3.5-turbo. This model demonstrated a high recall for identifying verifiable claims. However, it also exhibited a tendency to classify a significant number of genuinely verifiable claims as unverifiable, leading to a lower precision.

\textit{Second Pass: GPT-4}: To mitigate the precision issue and reduce the number of false negatives, we subjected all claims initially classified as 'Unverifiable' by GPT-3.5-turbo to a second round of evaluation using GPT-4. GPT-4, with its enhanced reasoning capabilities, provided a more nuanced assessment, reclassifying many of the false negatives from the first pass as 'Verifiable'.

\textit{Rationale for Two-Pass Approach}: This two-pass strategy allowed us to leverage the efficiency of GPT-3.5-turbo for initial screening while capitalizing on the superior accuracy of GPT-4 for refining the results. By focusing the computationally expensive GPT-4 only on the subset of claims flagged by GPT-3.5-turbo, we significantly reduced the overall cost of the verifiability check without compromising the quality of the final dataset.

For explicitly identifying claims that had associated media, we used the following prompt:
\begin{quote}
    For each of the following claims, determine if the claim explicitly depends on a specific piece of media (e.g., video, image, audio clip) to convey its full meaning. A claim should be considered 'media-associated' if it directly references or describes media content, such as "in the video," "as shown in the image," or "according to the audio clip." The claim should be marked as 'yes' only if understanding the claim relies on or is significantly enhanced by this media. If the claim makes sense on its own and does not require any media to be understood, respond with 'no.'
\end{quote}

For annotating the style of the claim, i.e., whether it was presented as a fact or expressed as an opinion, the following prompt was used.
\begin{quote}
    Determine the nature of each claim below, classifying it as either 'Factual' or 'Opinion', based on how it is framed in the text. Use these definitions:
    Factual: A claim asserted as a fact.
    Opinion: A claim that expresses personal belief or subjective interpretation.
    Please respond with the classification in English, even if the claim is in a non-English language. Analyze the content and context of each claim to determine its type.
\end{quote}
To validate the reliability of this classification, a random subset of 50 annotated claims was manually reviewed. The results were found to be accurate in 48 out of 50 cases. For example, factual claims such as \textit{“More than one-quarter of America’s young adults are too fat to serve in the U.S. military”} and \textit{“Alfamart Menyum-bangkan 6000 Kupon Untuk Membantu Melawan COVID-19”} (Translation: “Alfamart Announces Develop 6000 Coupons For Help Fighting COVID-19”) were correctly labeled as factual. In contrast, opinion claims like \textit{“Bernie Sanders’ health care plan would 'empower Republican governors to take away Medicaid, to take away health insurance for low-income and middle-income working Americans.'”} and \textit{“Se ele [João Doria] tivesse cumprido [a promessa de não deixar a prefeitura] não estaria com 60\% de rejeição na cidade de São Paulo”} (Translation: “If he [João Doria] had kept his promise not to leave the mayor's office, he wouldn't have 60\% disapproval in São Paulo”) were appropriately classified as opinions.

These examples highlight the model’s ability to generalize the classification task across both linguistic and cultural contexts, providing a reliable foundation for downstream analysis.

For extracting topics from the claims, the following prompt was used.
\begin{quote}
    Classify each of the following claims into one of the following categories: 'Health and Pandemics', 'Politics and Governance', 'Society and Culture', 'Economy and Environment', or, 'Conflict and Security'. Please respond with the category name in English, even if the claim is in a non-English language. Analyze the claim's content and determine the most relevant category.
\end{quote}

For extracting the presence of certain features in the claim, we used the following few-shot prompt:
\begin{quote}
    For each claim provided, identify and label the specific features it contains. The features to look for are: "Numerical Data", "Entity/Event Properties", "Position Statements", and "Quote". A claim may also have 'None' of these features. Use your pre-training knowledge to understand the claims based on their content and linguistic context. Here are examples to guide your judgment:

    Input:
    
    1. "People all over the world enjoy music." - This claim does not contain any of the specified features.
    
    2. "During his interview on March 3, 2021, the CEO stated, 'Our profits have doubled in the last two years due to our innovative approach.'" - This claim includes a "Quote" and "Numerical Data".
    
    3. "During the 2022 Climate Action Summit, the Canadian Minister of Environment declared that Canada will commit to zero emissions by 2050, aligning with the Paris Agreement goals" - This claim contains "Entity/Event Properties", and "Position Statements".

    Output:
    
    1. None
    2. Quote, Numerical Data
    3. Entity/Event Properties, Position Statements
\end{quote}

\paragraph{Fact-Checking Prompt (All Models)} 
We used a standardized base prompt to enable consistent fact-checking across models. This prompt was minimally adapted for certain architectures to accommodate interface-specific formatting or output behaviors. The objective was to elicit deterministic, label-only responses using the labels \texttt{true}, \texttt{false}, or \texttt{no verdict}.

\textbf{Base Prompt (Reference)} 
\begin{quote}
\small
\texttt{Fact check the following statements. Use labels ‘true’ or ‘false’. If a verdict cannot be reached, use ‘no verdict’. The output should only contain the numbered labels corresponding to each statement.}
\end{quote}

\textbf{Model-specific Prompt Variants} 

\begin{itemize}
    \item \textbf{GPT-3.5 Turbo and GPT-4 (OpenAI API)}\\
    \textit{System Prompt:} Default\\
    \textit{User Prompt:} Used the base prompt without any changes.

    \item \textbf{LLaMA 3.1 8B (local server) and LLaMA 3.1 70B (Groq API)}\\
    \textit{System Prompt:}
    \begin{quote}
    \small
    \texttt{You are a multilingual language model with pretraining knowledge about the world.}
    \end{quote}
    \textit{User Prompt:}
    \begin{quote}
    \small
    \texttt{Fact check the following statements. Use labels ‘true’ or ‘false’. If a verdict cannot be reached, use ‘no verdict’. Output should only contain the assigned labels.}
    \end{quote}

    \item \textbf{Mixtral 8x7B (Groq API)}\\
    \textit{System Prompt:}
    \begin{quote}
    \small
    \texttt{You are a multilingual language model with pretraining knowledge about the world.}
    \end{quote}
    \textit{User Prompt:}
    \begin{quote}
    \small
    \texttt{Fact check the following statements. For each statement, provide the assigned label in the format '1. True', '2. False', etc. Use 'true' or 'false' as labels. If a verdict cannot be reached, use 'no verdict'. The output should only contain the numbered labels in English corresponding to each statement.}
    \end{quote}
\end{itemize}

Prompt adaptations were developed through iterative manual refinement on a small set of test claims. Beginning with a simple base prompt, we introduced minor modifications until each model produced consistent and reliable outputs aligned with the expected labeling format.

\subsection*{Model Selection and Fact-Checking}

\subsubsection*{Models and Hyperparameters}
The following models were used in our experiments:
\begin{itemize}
    \item \textbf{GPT-4o:} \textit{gpt-4o-2024-05-13} accessed via OpenAI API.
    \item \textbf{GPT3.5} \textit{gpt-3.5-turbo-0125} accessed via OpenAI API.
    \item \textbf{LLaMA 3.1 8B:} \textit{meta-llama/Meta-Llama-3.1-8B}, downloaded from HuggingFace, locally deployed on GPU.
    \item \textbf{LLaMA 3.1 70B:} \textit{llama-3.1-70b-versatile}, accessed via Groq API.
    \item \textbf{Mixtral 8x7B:} \textit{mixtral-8x7b-32768}, accessed via Groq API.
\end{itemize}
The speed of processing varied significantly across models. The slowest was the locally deployed LLaMA 3.1 8B, which processed approximately 5 claims every 12 seconds. The fastest was GPT-4o, which processed approximately 30 claims every 2 seconds.

\subsubsection*{Language distribution}
The detailed distribution of languages in the dataset is given in Table \ref{tab:lang_dist}.

\begin{table}[htbp]\centering
\caption{Language Distribution of the Dataset}
\label{tab:lang_dist}
\begin{tabular}{|l|c|}
\hline
\textbf{Language} & \textbf{Count} \\
\hline
English & 23215 \\
Portuguese & 8371 \\
Spanish & 3497 \\
German & 2294 \\
Indonesian & 2160 \\
Tamil & 2078 \\
Hindi & 1918 \\
Arabic & 1857 \\
Turkish & 1754 \\
Polish & 1660 \\
Italian & 1523 \\
Telugu & 1445 \\
Dutch & 1043 \\
Romanian & 913 \\
French & 844 \\
Persian & 833 \\
Serbian & 804 \\
Bengali & 774 \\
Georgian & 770 \\
Sinhala & 659 \\
Russian & 573 \\
Albanian & 456 \\
Norwegian & 338 \\
Kannada & 318 \\
Filipino & 315 \\
Hebrew & 303 \\
Chinese & 280 \\
Azerbaijani & 227 \\
Assamese & 173 \\
Khmer & 118 \\
\hline
\end{tabular}
\end{table}

\subsubsection*{Language Classification Details}
To provide detailed insight into the language resource availability, we leveraged the language resource taxonomy developed by Joshi et al. \cite{joshi_state_2020}. This taxonomy, which categorizes languages into six classes (0-5) based on digital resource richness, is maintained on GitHub at \url{https://microsoft.github.io/linguisticdiversity/}.

We utilized the \texttt{lang2tax.txt} file from the repository (\url{https://microsoft.github.io/linguisticdiversity/assets/lang2tax.txt}) for language-to-class mapping. While most languages were directly mapped, we implemented the following specific adjustments:

\begin{itemize}
    \item Norwegian: Mapped to Class 1, considering the separate listings for Norwegian (bokmål) and Norwegian (nynorsk), both in Class 1.
    \item Filipino: Assigned the Class value of Tagalog.
    \item Chinese: Assigned the Class value of Mandarin.
\end{itemize}

This detailed language classification provides granular data for analyzing LLM performance across varied linguistic resource contexts.
\begin{figure}
\centering
\includegraphics[width=\textwidth]{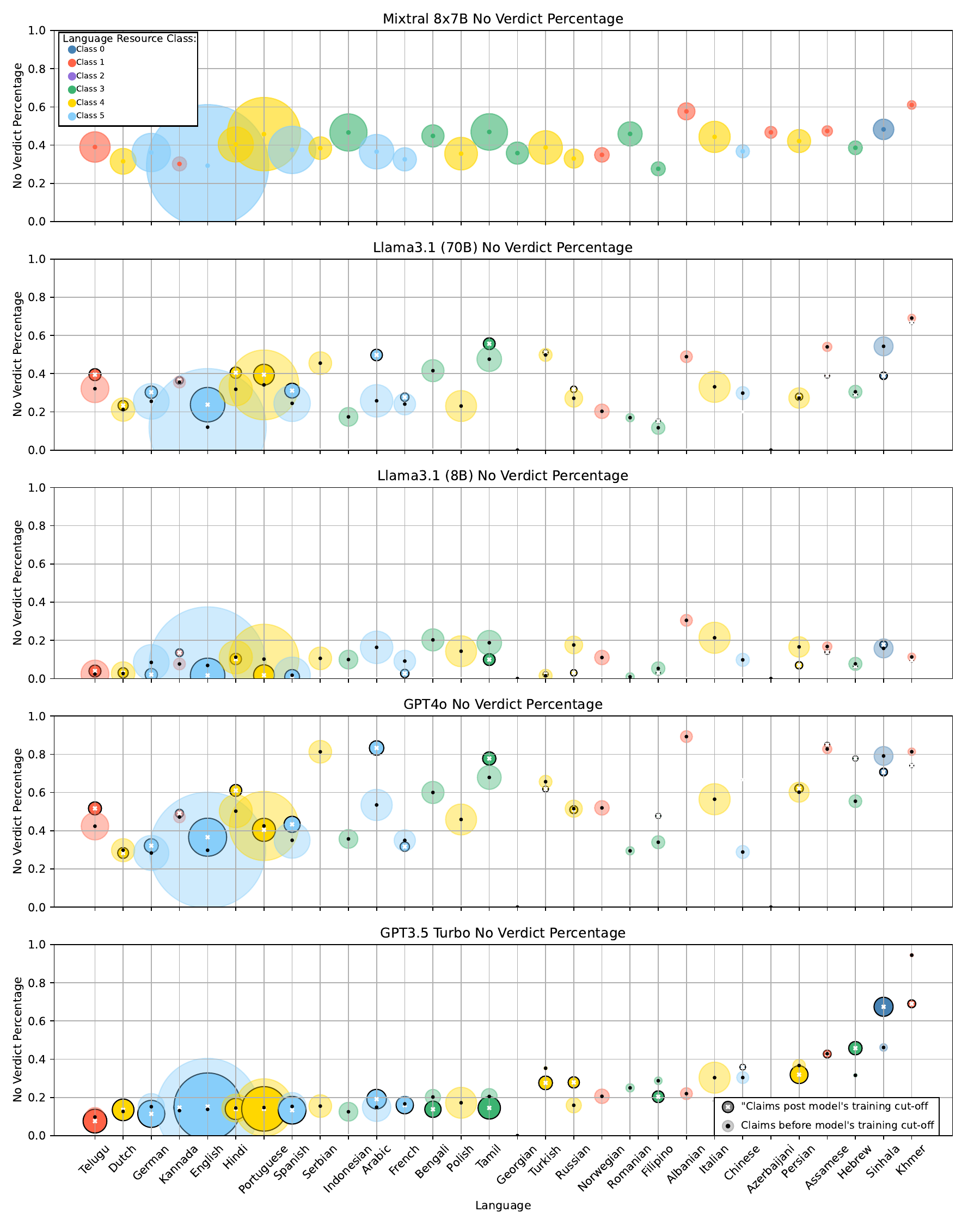}
\caption{\footnotesize No verdict percentages of five LLMs (Mixtral 8x7B, LLaMA 3.1 (70B and 8B), GPT-4o, and GPT-3.5 Turbo) across 61,523 claims in multiple languages. Each subplot represents one model, with the x-axis showing different languages and the y-axis representing the no verdict percentages. For each language, we display the model's performance on claims created before the model’s training cut-off (black dot) and after the cut-off (white cross). The size of the circle around each marker corresponds to the volume of claims in that language. The figure highlights the variation in the fact checking coverage across languages and between models, showing that the best performing GPT4o model also had the highest reservation while choosing claims to fact check.}
\label{fig:refusal_rates}
\end{figure}

\section*{Results}

\subsection*{Fact-Checking Performance of LLMs}

\subsubsection*{Model Refusal Rates}
The no-verdict percentage of each model for each language tested is shown in Figure \ref{fig:refusal_rates}. 

\subsubsection*{Example Model Verdicts}

The table \ref{tab:model_verdicts} presents a small snippet of fact-checking results from the five Large Language Models (LLMs) we evaluated: GPT-4o, GPT-3.5 Turbo, LLaMA 3.1 8B, LLaMA 3.1 70B, and Mixtral 8x7B. It showcases the models' verdicts on a selection of claims, along with the corresponding language and the fact-checkers' ground truth label.

\subsection*{Pre-Cutoff vs. Post-Cutoff Performance}

In the main paper, we evaluated the models' performance on claims created before and after their respective training data cut-off points, observing that closed-source models, particularly GPT-3.5 Turbo and GPT-4o, demonstrated robust generalization on post-cutoff data. However, as the study was conducted on claims up to July 2024, an additional analysis was performed to further explore model performance on more recent data, specifically claims made after this date.

To this end, a new dataset was compiled, containing 3,033 claims ranging from August 1, 2024, to February 19, 2025, after standardizing labels and filtering for validity. The models' refusal rates on this new dataset were notably higher for closed-source models compared to the initial study: GPT-4o at 51.6\% (previously 43.02\%) and GPT-3.5 Turbo at 34.7\% (previously 16.57\%). This increase in refusal rates, especially for GPT-3.5 Turbo, may indicate the models' recognition of their knowledge limitations for claims requiring post-training data knowledge. In contrast, open-source models (LLaMA and Mixtral) showed more consistent refusal rates.

To refine the accuracy assessment, claims were annotated for date relevance using GPT-4o, identifying those explicitly referencing events post-cutoff. This allowed for a focused accuracy evaluation on date-relevant claims, yielding the following results: GPT-3.5 Turbo at 74.55\% (previously 79.75\%), GPT-4o at 83.5\% (previously 80.45\%), LLaMA 3.3 (70B) Groq at 62.5\%, LLaMA 3.1 (8B) Local (Quantized) at 57.6\% (previously 68.55\%), LLaMA 3.1 (8B) Groq at 61.1\%, and Mixtral at 62.5\%. Notably, GPT-4o's accuracy increased slightly, though it only provided verdicts for 917 out of the 1970 date-relevant claims.

These results reinforce the observation that closed-source models, particularly GPT-4o, maintain strong performance on post-training data, even after filtering for date-relevant claims. Furthermore, the analysis revealed interesting variations in LLaMA 3.1 8B's performance between Groq API and local quantized instances, with the latter showing significantly lower refusal rates but also slightly lower accuracy. This highlights the impact of model deployment and quantization on performance.

The annotation prompt used to determine date relevance was tailored to each model's cut-off date. The following prompt, for example, was used to identify claims relevant to GPT-3.5 Turbo’s cut-off date of September 2021:

\begin{quote}
"For each of the following claims, determine if the claim is about a specific event that took place after September 2021.
A claim should be considered 'yes' if it explicitly refers to an event that occurred after this date, making it impossible to fact-check without post- September 2021 knowledge. Examples include references to recent elections, policy changes, newly published scientific findings, or breaking news events.
If the claim is general (e.g., "The Earth is round") or could be fact-checked using pre-September 2021 knowledge, respond with 'no'." 
\end{quote}
-- \textit{used with temperature = 0}

Analogous prompts were crafted for the other models, substituting their respective cut-off dates.

Despite expectations that non-date-relevant claims would exhibit higher accuracy, the difference between date-relevant and non-date-relevant claim accuracy was minimal. This observation necessitates further investigation to fully understand the factors influencing model performance on post-cutoff data.

Table \ref{tab:date_relevance} illustrates the accuracy of each model for both date-relevant and non-date-relevant claims. It highlights the consistency of closed-source model performance, particularly GPT-4o, and the variability observed in open-source models, especially LLaMA 3.1 8B across different deployment modes.

\begin{table}[htbp]\centering
\caption{Model Accuracy on Date relevant and Non-Date relevant claims}
\label{tab:date_relevance}
\begin{tabular}{|p{4cm}|p{4cm}|p{4cm}|}
\hline
\textbf{Model} & \textbf{Date-Relevant Accuracy(\%)} & \textbf{Non Date-Relevant Accuracy(\%)} \\
\hline
GPT3.5 Turbo & 74.5565 & 75.1445 \\ \hline
GPT4o & 83.5332 & 81.7028 \\ \hline
Llama3.3 70B (Groq API) & 62.4839 & 65.1162 \\ \hline
Llama3.1 8B (Local) & 57.5924 & 56.1428 \\ \hline
Llama3.1 8B (Groq API) & 61.0906 & 63.4768 \\ \hline
\end{tabular}
\end{table}

In conclusion, while the closed-source models continue to demonstrate superior post-cutoff performance, open-source models require further refinement. Future work should explore datasets specifically designed to test post-cutoff knowledge, such as claims related to recent elections or newly unfolding events, to provide a more robust evaluation.

\begin{table}[htbp]\centering
\caption{Example Model Verdicts.}
\label{tab:model_verdicts}
\small
\begin{tabular}{|p{4.0cm}|p{1.5cm}|p{2.5cm}|p{1.3cm}|p{1.3cm}|p{1.3cm}|p{1.3cm}|p{1.3cm}|}
\hline
\textbf{Claim Text} & \textbf{Language} & \textbf{Fact Checkers' Verdict} & \textbf{GPT-4o} & \textbf{GPT-3.5 Turbo} & \textbf{LLaMA 3.1 8B} & \textbf{LLaMA 3.1 70B} & \textbf{Mixtral 8x7B} \\
\hline 
Son 50 yılın siyasi tarihine bakın, tek başına iktidarlardaki Türkiye’nin büyüme oranı, koalisyon dönemlerindekinin iki katıdır. & tr & False & No verdict & True & True & True & True \\ \hline
Aqui em São Paulo, nós estamos com 2,7 milhões de pessoas vivendo com os tais R\$ 600 que viraram R\$ 300 [do auxílio emergencial] & pt & Partly True/Misleading & No verdict & False & True & True & True \\ \hline
Das Coronavirus enthält HIV-Anteile, wurde also im Labor erschaffen & de & False & False & False & False & False & False \\ \hline
„U prvim mesecima 2011. godine imamo odlične pokazatelje izvoza poljoprivrednih proizvoda iz Srbije, koji je za 40 odsto veći nego prošle godine.“ & sr & True & No verdict & False & False & No verdict & True \\ \hline
Quanto spendono Italia ed Europa per l’immigrazione & it & Mostly True & No verdict & No verdict & No verdict & No verdict & No verdict \\ \hline
Alfamart Menyumbangkan 6000 Kupon Untuk Membantu Melawan COVID-19 & id & False & No verdict & True & True & True & True \\ \hline
\end{tabular}
\end{table}

\section*{Materials}

\subsection*{Dataset Snippets}

\subsubsection*{Base Dataset}

Table \ref{tab:base_dataset} presents the core dataset, constructed from web-sourced claims identified via ClaimReview Markup and validated against reputable fact-checking organizations. This dataset comprises essential information extracted to facilitate fact-checking analysis, providing a clean, fundamental view of the claims.

\begin{table}[htbp]\centering
\caption{Snippet of the Dataset.}
\label{tab:base_dataset}
\small 
\begin{tabular}{|p{7.5cm}|p{1.5cm}|p{2cm}|p{2.5cm}|} 
\hline
\textbf{Claim Text} & \textbf{Language} & \textbf{Claim Date} & \textbf{Fact Checkers' Verdict} \\
\hline
Son 50 yılın siyasi tarihine bakın, tek başına iktidarlardaki Türkiye’nin büyüme oranı, koalisyon dönemlerindekinin iki katıdır. & tr & none & False \\ \hline
Aqui em São Paulo, nós estamos com 2,7 milhões de pessoas vivendo com os tais R\$ 600 que viraram R\$ 300 [do auxílio emergencial] & pt & 2020-11-03T12:41:12Z & Partly True/Misleading \\ \hline
Das Coronavirus enthält HIV-Anteile, wurde also im Labor erschaffen & de & 2020-04-17T23:00:00Z & False \\ \hline
„U prvim mesecima 2011. godine imamo odlične pokazatelje izvoza poljoprivrednih proizvoda iz Srbije, koji je za 40 odsto veći nego prošle godine.“ & sr & 2011-05-14T00:00:00Z & True \\ \hline
Quanto spendono Italia ed Europa per l’immigrazione & it & 2017-08-02T22:00:00Z & Mostly True \\ \hline
Alfamart Menyumbangkan 6000 Kupon Untuk Membantu Melawan COVID-19 & id & none & False \\
\hline
\end{tabular}
\end{table}

\subsubsection*{Annotated Dataset} Table \ref{tab:annotated_dataset} illustrates the enriched version of the base dataset. This augmented dataset incorporates additional annotations derived through a combination of data wrangling techniques and Large Language Model (LLM) assistance, offering a more comprehensive and detailed perspective on each claim's characteristics.
\begin{table}[htbp]\centering
\caption{Snippet of the Annotated Dataset.}
\label{tab:annotated_dataset}
\footnotesize
\setlength{\tabcolsep}{3pt}
\begin{tabular}{|p{2.0cm}|p{0.9cm}|p{1.6cm}|p{1.6cm}|p{0.82cm}|p{1.2cm}|p{1.2cm}|p{0.82cm}|p{0.99cm}|p{1.1cm}|p{1.3cm}|p{0.82cm}|p{1.1cm}|} 
\hline
\textbf{Claim Text} & \makecell{\textbf{Lang-}\\\textbf{uage}} & \textbf{Claim Date} & \textbf{Fact Checkers' Verdict} & \textbf{Claim Year} & \textbf{Position Statements} & \makecell{\textbf{Entity/}\\\textbf{Event}\\\textbf{Propert-}\\\textbf{ies}} & \textbf{Quote} & \makecell{\textbf{Numer-}\\\textbf{ical}\\\textbf{Data}} & \textbf{Claim Type} & \textbf{Topics} & \textbf{Media Associated} & \textbf{Mapped Label} \\
\hline
Son 50 yılın siyasi tarihine bakın, tek başına iktidarlardaki Türkiye’nin büyüme oranı, koalisyon dönemlerindekinin iki katıdır. & tr & none & False & & 1.0 & 0.0 & 0.0 & 0.0 & Factual & Politics and Governance & No & False \\ \hline
Aqui em São Paulo, nós estamos com 2,7 milhões de pessoas vivendo com os tais R\$ 600 que viraram R\$ 300 [do auxílio emergencial] & pt & 2020-11-03T12:41:12Z & \makecell{Partly\\True/\\Misleading} & 2020.0 & 0.0 & 0.0 & 0.0 & 1.0 & Factual & Economy and Environment & No & False \\ \hline
Das Coronavirus enthält HIV-Anteile, wurde also im Labor erschaffen & de & 2020-04-17T23:00:00Z & False & 2020.0 & 1.0 & 0.0 & 0.0 & 0.0 & Opinion & Health and Pandemics & No & False \\ \hline
„U prvim mesecima 2011. godine imamo odlične pokazatelje izvoza poljoprivrednih proizvoda iz Srbije, koji je za 40 odsto veći nego prošle godine.“ & sr & 2011-05-14T00:00:00Z & True & 2011.0 & 0.0 & 0.0 & 0.0 & 1.0 & Factual & Economy and Environment & No & True \\ \hline
Quanto spendono Italia ed Europa per l’immigrazione & it & 2017-08-02T22:00:00Z & Mostly True & 2017.0 & 0.0 & 0.0 & 0.0 & 1.0 & Opinion & Conflict and Security & No & True \\ \hline
Alfamart Menyumbangkan 6000 Kupon Untuk Membantu Melawan COVID-19 & id & none & False & & 0.0 & 1.0 & 0.0 & 1.0 & Factual & Society and Culture & No & False \\
\hline
\end{tabular}
\end{table}

\clearpage

\end{document}